\documentclass{ws-procs975x65}            
\begin{document}
\title{Black hole solutions in Einstein-charged scalar field theory}

\author{SUPAKCHAI PONGLERTSAKUL$^*$, SAM DOLAN and ELIZABETH WINSTANLEY}

\address{Consortium for Fundamental Physics, School of Mathematics and Statistics, \\ The University of Sheffield, Hicks Building, Hounsfield Road, \\
Sheffield, S3 7RH, United Kingdom\\
$^*$E-mail: smp12sp@sheffield.ac.uk \\
}

%

\begin{abstract}
We investigate possible end-points of the superradiant instability for a charged black hole with a reflecting mirror. By considering a fully coupled system of gravity and a charged scalar field, hairy black hole solutions are obtained. The linear stability of these black hole solutions is studied.
\end{abstract}


\bodymatter

\section{Introduction}\label{sp:sec1}
In classical general relativity, there is a process by which energy can be extracted from a black hole. This phenomenon is known as superradiant scattering. More specifically, the amplitude of a bosonic field around black hole will be amplified if the frequency $\omega$ and the charge $q$ of the field satisfy \cite{Bekenstein:1973mi}, $\omega < q \Phi_{\mathrm{H}}$,
where $\Phi_{\mathrm{H}}$ is electric potential at the outer horizon. Note that this inequality is the charged version of superradiant condition.

Via superradiant scattering, an instability of the background spacetime can occur if the bosonic field can be trapped within the proximity of the black hole. Wave modes are repeatedly scattered off and amplified by a black hole, so ultimately their back-reaction on the background becomes non-negligible. For a charged scalar field with non-zero mass, this instability can be triggered by surrounding a charged black hole with a reflecting mirror \cite{Herdeiro:2013}. 




Superradiant instabilities for charged black holes have been extensively studied in the linear regime. Although, the end-point of these instabilities remains unknown and a fully non-linear analysis is needed. Therefore in this talk, one possible end-points of the superradiant instability for charged black holes in a cavity is investigated. When a fully coupled system of gravity and a charged scalar field with a mirror is considered, we find black hole solutions with scalar field hair. Considering linearised perturbations of these black holes, we perform a numerical analysis of the stability of these hairy black hole solutions. Here we present a summary of our results. More details of this work are to appear in Ref.~\citenum{SP:2015}.


\section{Massless charged scalar perturbation of Reissner-Nordstr\"{o}m spacetime with a mirror}\label{sp:sec2}
As a preliminary, we consider a massless charged scalar field $\phi$ on the Reissner-Nordstr\"{o}m (RN) background with a mirror. In the test-field approximation, the usual Klein-Gordon (KG) equation in curved spacetime is obtained. This can be solved by using the ansatz $\phi\sim e^{-i\sigma t}R(r)$ where $\sigma$ and $R(r)$ are respectively the frequency and the radial part of the scalar field. We impose an ingoing wave condition near the horizon $r\rightarrow r_{\mathrm{h}}$ and at the mirror the scalar field vanishes $R(r_{\mathrm{m}})=0$, where $r_{\mathrm{m}}$ is the location of the mirror. We integrate the KG equation numerically and obtain the corresponding frequency. Figure~\ref{sp:fig1} shows an example plot of the frequency of the scalar field against the mirror radius when the black hole mass is fixed to be $M=1$. The superradiant instability can be clearly seen from Figs.~\ref{sp:fig1}(a) and \ref{sp:fig1}(b), since there are regions where Im$(\sigma)>0$ indicating an unstable mode.


\def\figsubcap#1{\par\noindent\centering\footnotesize(#1)}
\begin{figure}[h]%
\begin{center}
  \parbox{2.1in}{\includegraphics[width=2in]{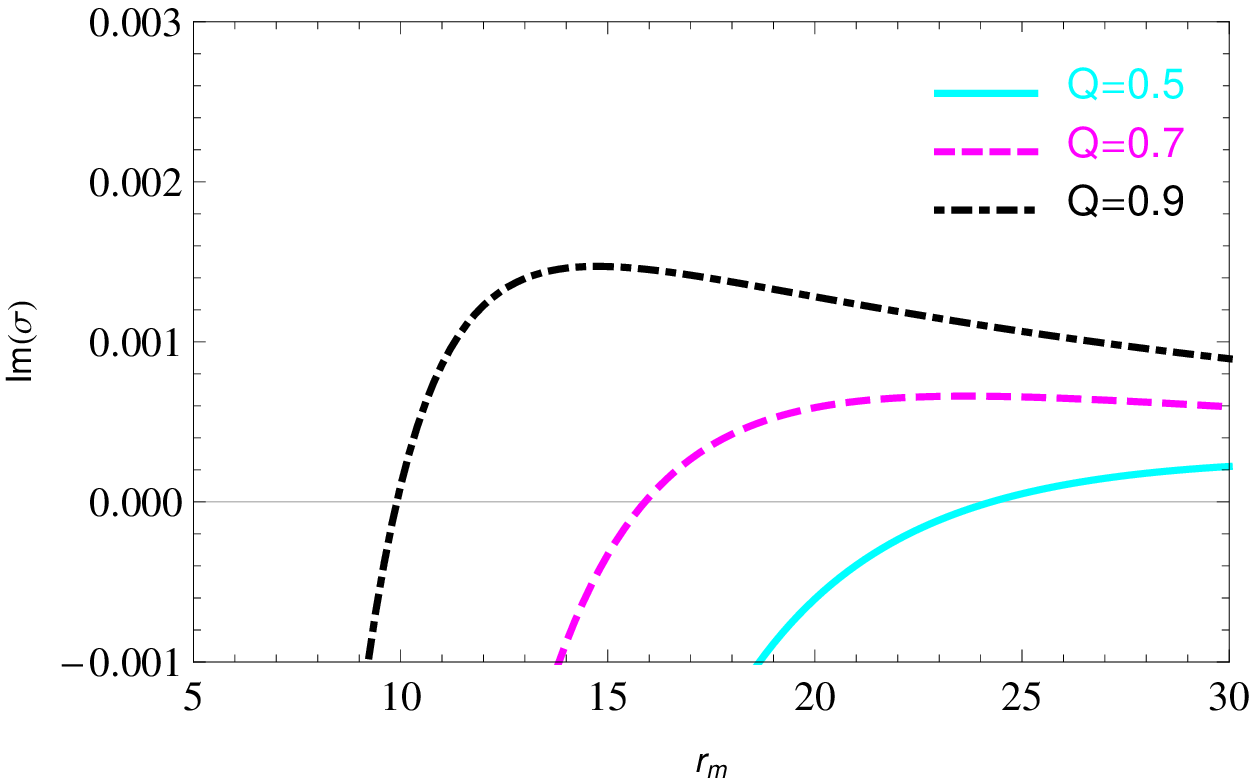}
  \figsubcap{a}}
  \hspace*{4pt}
  \parbox{2.1in}{\includegraphics[width=2in]{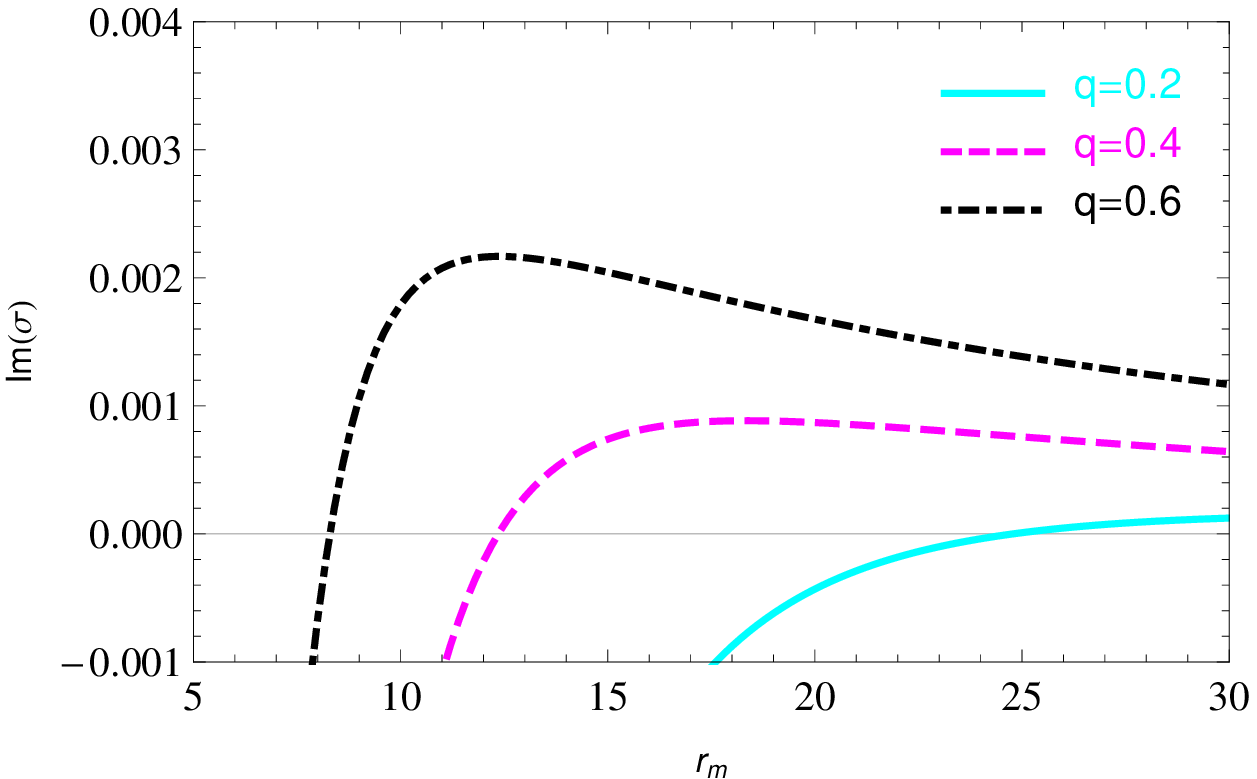}
  \figsubcap{b}}
  \caption{The imaginary part of $\sigma$ is plotted as a function of the location of the mirror $r_{\mathrm{m}}$, (a) for fixed scalar charge $q=0.5$ and different values of the black hole charge $Q$, (b) for fixed $Q=0.9$ and different values of $q$.}%
  \label{sp:fig1}
\end{center}
\end{figure}

Therefore from Fig.~\ref{sp:fig1}, a massless charged scalar field on the RN spacetime with a mirror experiences a superradiant instability. These results are in agreement with with previous work done by Herdeiro {\it et al.}\cite{Herdeiro:2013}, where a superradiant instability of a massive charged scalar field on a charged black hole in a cavity has been discussed. To proceed further, in the next section, we will investigate a fully coupled system of gravity and a charged scalar field.

\section{Hairy black hole in Einstein-Maxwell-Klein-Gordon theory with a mirror}\label{sp:sec3}
We are interested in a fully coupled system of gravity and a charged scalar field. This system is described by the action
\begin{equation}\label{sp:eq2}
S  = \int d^{4}x \sqrt{-g} \left[ \frac{R}{16\pi G} - \frac{1}{4} F_{ab}F^{ab} -\frac{1}{2}g^{ab} D^\ast_{(a} \phi^\ast D^{}_{b)} \phi \right],
\end{equation}
where $F_{ab} = \nabla_a A_b - \nabla_b A_a$, $A_a$ is the electromagnetic potential, $D_a = \nabla_a - iqA_a$, $q$ is the scalar field charge and $X_{(ab)}=\frac{1}{2}\left(X_{ab}+X_{ba}\right)$.
The line element of a static spherically symmetric spacetime is given by
\begin{equation}
ds^2 = -f(r)h(r)dt^2 + f(r)^{-1}dr^2 + r^2\left(d\theta^2 + \sin^2\theta~d\varphi^2\right),\label{sp:eq6}
\end{equation}
the electromagnetic vector potential is $A_a\equiv[A_{0}(r),0,0,0]$ and the scalar depends on $r$ only $\phi=\phi(r)$. Varying Eq.~(\ref{sp:eq2}), three equations of motion are obtained, which can be solved numerically.

\begin{figure}[h]%
\begin{center}
  \parbox{2.1in}{\includegraphics[width=2in]{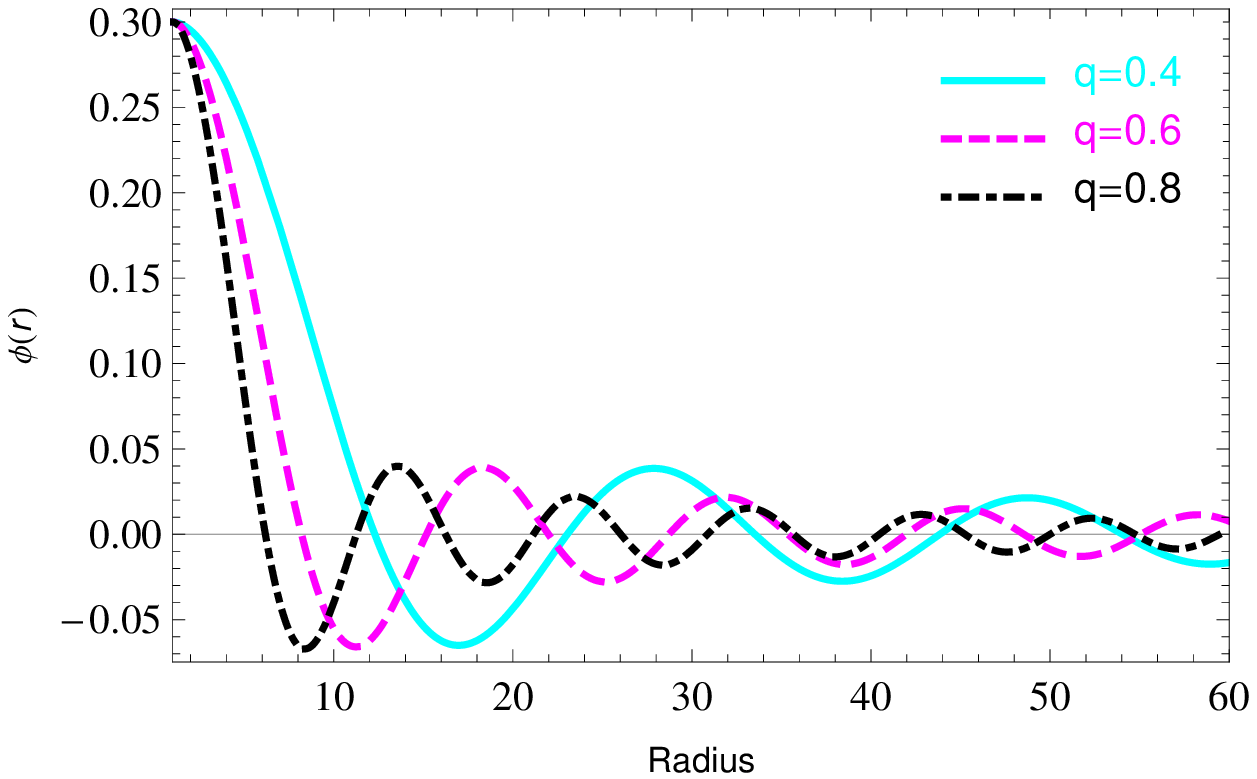}
  \figsubcap{a}}
  \hspace*{4pt}
  \parbox{2.1in}{\includegraphics[width=2in]{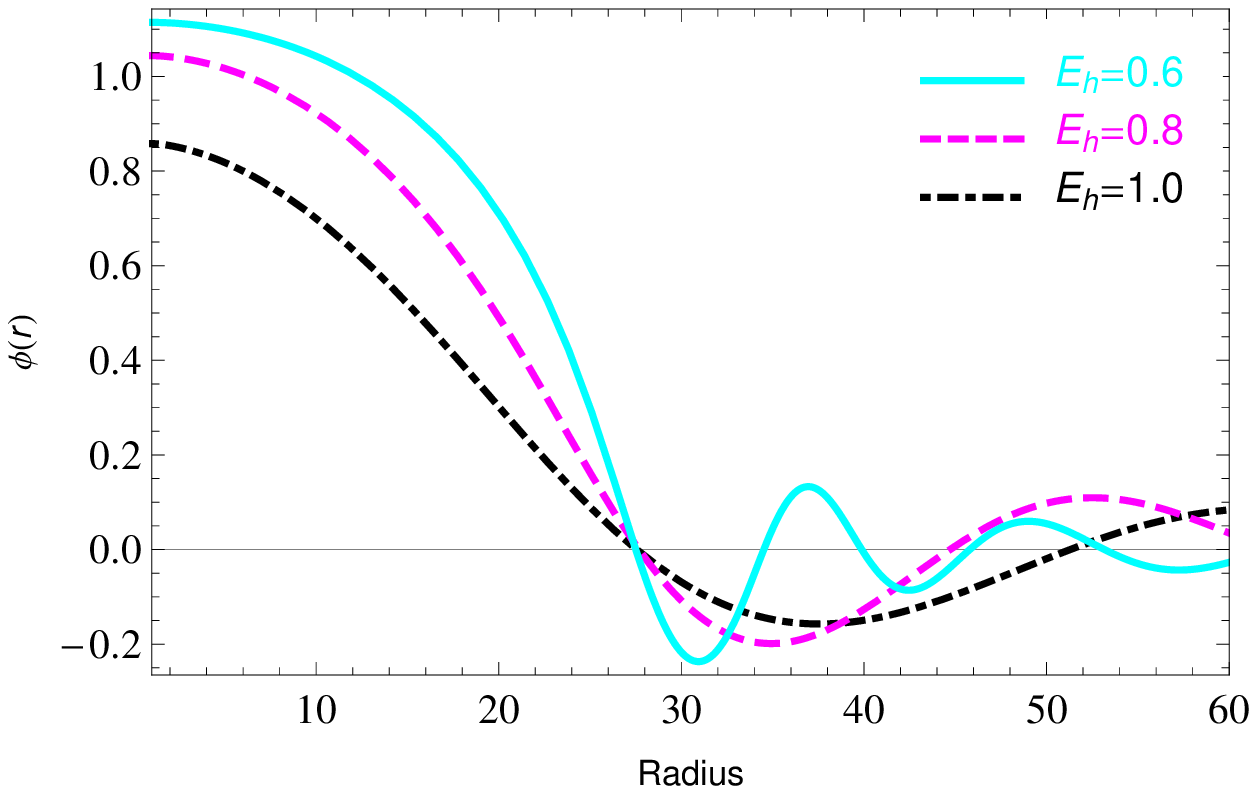}
  \figsubcap{b}}
  \caption{The scalar field $\phi(r)$ is plotted as a function of radius (a) for fixed $\phi_{\mathrm{h}}=0.3$, $E_{\mathrm{h}}=0.6$ and different values of $q$, (b) for fixed $q=0.1$ and different values of $E_h$. }%
  \label{sp:fig2}
\end{center}
\end{figure}

In Fig.~\ref{sp:fig2}, the profile of the scalar field outside the black hole event horizon is illustrated, where the black hole radius is fixed at $r_{\mathrm{h}}=1$. For each curve, there are three parameters that need to be specified, namely, $\phi_{\mathrm{h}}$ the value of the scalar field on the horizon, $E_{\mathrm{h}} \equiv A'_{0}(r_{\mathrm{h}})$ the electric field on the horizon and $q$. The oscillatory behaviour of the scalar field allows us to put the mirror at any zero of $\phi$. Here we consider only the case for which the mirror is located at the first zero of $\phi$. This is because these black hole solutions with the mirror at the first zero are expected to be stable.

By changing one of these three parameters, we obtain different hairy black hole solutions. Figure~\ref{sp:fig2}(a) displays three distinct black hole solutions with varying scalar charge and three different mirror radii. However, as shown in Fig.~\ref{sp:fig2}(b), it is possible that different solutions could share the same mirror location.

\section{Stability of the hairy black holes}\label{sp:sec4}
We are now ready to perform a stability analysis of these equilibrium solutions. If these solutions are shown to be stable, they could represent the end-point of the superradiant instability for a massless charged scalar perturbation on the RN background with a mirror as discussed in Sec.~\ref{sp:sec2}. 

To investigate the stability of the equilibrium solutions, we perturb the four field variables $(f,h,A_{0},\phi)$ in the theory as follows, $f=\bar{f}(r) + \delta f(t,r)$ and similarly for the other three quantities. In this formalism, $\bar{f}$ is the equilibrium quantity and $\delta f$ is the perturbation. By linearising the field equations, three coupled perturbation equations for $\delta A_0$ and the real and imaginary parts of $\delta\phi$ are obtained. The perturbation equations are solved by assuming $\delta\phi\sim e^{-i\sigma t}\tilde{\phi}(r)$, and similarly for the other perturbations. These perturbation modes need to satisfy the following boundary conditions, $\tilde{\phi}(r)$ and other perturbations modes follow an ingoing wave-like condition near the horizon. The scalar field vanishes at the mirror $(\tilde{\phi}(r_{\mathrm{m}})=0)$.

\begin{figure}[h]%
\begin{center}
  \parbox{2.1in}{\includegraphics[width=2in]{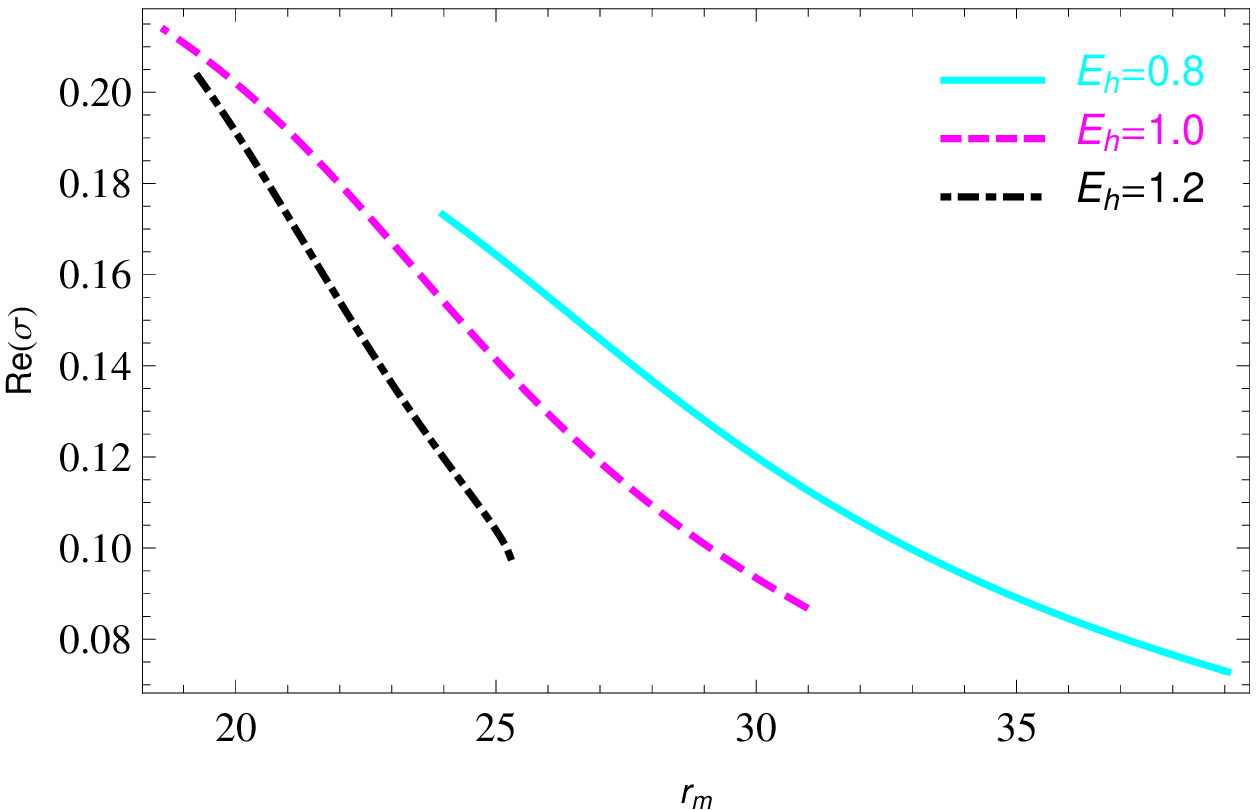}
  \figsubcap{a}}
  \hspace*{4pt}
  \parbox{2.1in}{\includegraphics[width=2in]{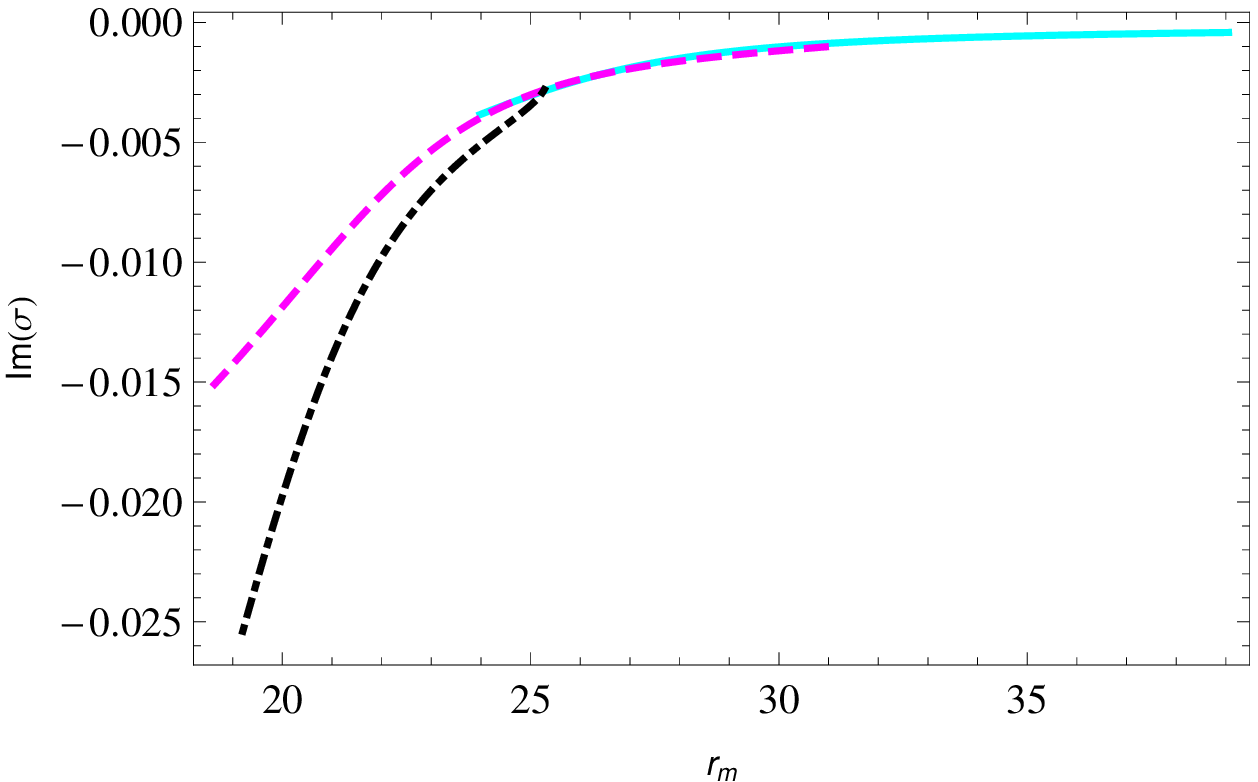}
  \figsubcap{b}}
  \caption{The real (a) and imaginary part (b) of the perturbation frequency $\sigma$ are plotted against the location of the mirror $r_{\mathrm{m}}$ with the scalar charge is fixed at $q=0.1$, $\phi_{\mathrm{h}}$ varying from $0.1$ to $1.4$ and different values of $E_h$.}%
  \label{sp:fig3}
\end{center}
\end{figure}
We numerically integrate the coupled perturbation equations by scanning for a frequency $\sigma$ such that the boundary conditions are satisfied. An example plot is shown in Fig.~\ref{sp:fig3}. In each plot, the charge of the scalar field is $q=0.1$ and the initial value of the scalar field varies between $\phi_{\mathrm{h}}=0.1-1.4.$ Each point in this plot represents the perturbation frequency for one distinct hairy black hole solution. More precisely, with the mirror at the first zero of the static scalar field $\phi$, we find one value of $\sigma$ that satisfies the boundary conditions. The key result here is in Fig.~\ref{sp:fig3}(b) where Im$(\sigma)<0$ indicates that the perturbation modes decay exponentially in time. Thus it is clearly shown that these black hole solutions are linearly stable. 

\section{Summary}\label{sp:sec5}
We have studied the possible end-point of the superradiant instability of a massless charged scalar field on the RN spacetime with a mirror. Numerical black hole solutions with charged scalar hair are obtained. Numerical study suggests that these hairy black hole solutions appear to be stable. Thus, they could represent the end-point of the superradiant instability.


\section*{Acknowledgments}
The work of EW and SRD is supported by the Lancaster-Manchester-Sheffield Consortium for Fundamental Physics under STFC grant ST/L000520/1. The work of SRD is supported by EPSRC grant EP/M025802/1.




\end{document}